\documentclass[12pt]{article}
\usepackage{amsmath}
\usepackage{graphicx}
\newcommand{\la}{\langle}
\newcommand{\ra}{\rangle}
\newcommand{\Tr}{{\rm Tr}}
\renewcommand{\Im}{{\rm Im}\,}
\newcommand{\be}{\begin{equation}}
\newcommand{\ee}{\end{equation}}
\newcommand{\bea}{\begin{eqnarray}}
\newcommand{\eea}{\end{eqnarray}}
\newcommand{\nl}{\nonumber \\}

\newcommand{\psibar}{\,\overline{\phantom{I}}\!\!\!\!\psi}
\newcommand{\Dslash}{\,/\!\!\!\!D}

\newcommand{\pslash}{\,/\!\!\!p}
\newcommand{\vslash}{\,/\!\!\!v}

\newcommand{\Eq}[1]{{\rm Eq.\,}(\ref{#1})}
\newcommand{\eq}[1]{{\rm eq.\,}(\ref{#1})}
\newcommand{\eqs}[1]{{\rm eqs.\,}(\ref{#1})}
\newcommand{\refeq}[1]{(\ref{#1})}
\newcommand{\OO}{{\cal O}}

\newcommand{\lsim}{\stackrel{<}{{}_{\sim}}}

\begin{document}
\baselineskip=15.5pt
\pagestyle{plain}
\setcounter{page}{1}

\begin{titlepage}


\vskip -.8cm


\begin{center}

\vskip 1.7 cm

{\LARGE {On Supersymmetry at Finite Temperature}}

\vskip 1.5cm
{\large 
Simon Caron-Huot$^{\dagger}$}

\vskip 1.2cm

$^{\dagger}$ Department of physics, McGill University,
Montr\'eal, Canada.

{\tt scaronhuot@physics.mcgill.ca}

\medskip
\vskip .5cm

\vskip 1.7cm

{\bf Abstract}
\end{center}

\noindent

We consider the effective theories governing the
sensitivity to the plasma
of certain high-energy observables in supersymmetric plasmas,
and point out that they preserve supersymmetry.
Our findings generalize previous observations on asymptotic
thermal masses in weakly coupled plasmas, to both the real and imaginary parts of
self-energies, on the light cone and away from it,
in weakly and strongly interacting theories.
All observed supersymmetry violations due to thermal effects
turn out to vanish faster than $E^{-2}$ in the high energy limit.

\end{titlepage}

\newpage

\section{Introduction}

Supersymmetry is usually considered as being broken at finite
temperature (see e.g. \cite{girardello80}).
A simple reason for this is the different
statistics and population functions assumed by Bose and Fermi fields,
making it hard to see how a Bose-Fermi symmetry could be preserved.
In Euclidean space, Bose-Fermi symmetry is explicitly broken
by boundary conditions along the periodic time direction
(with period $1/T$), which are respectively periodic for bosonic fields
and anti-periodic for fermionic fields.
Nevertheless, one can still ask whether the supersymmetry
of the underlying equations of motion leaves any trace in physical
observables.

One implication of supersymmetry at finite temperature along these lines
was described in \cite{yaffehydro}:
due to the existence of a conserved supercurrent, the effective
hydrodynamics theory
which describes the long-wavelength modes of the plasma must contain
fermionic degrees of freedom.
These fermionic modes enter the discussion of non-linear effects within this
effective theory (e.g. loop corrections),
but, of course, are not allowed to take on classical expectation values.

In this letter we propose to look at a different
sector of the theory: that of high-energy observables.
In the strict high-energy limit the plasma decouples and
supersymmetry is recovered (provided it is present in the vacuum theory).
Nontrivial results may be obtained by looking at
the leading thermal corrections to this limit,
such as thermal-induced dispersion relations and decay rates.
We see no obvious reason why these should preserve supersymmetry.
Nevertheless, in this letter we wish to report an
intriguing fact: for a wide class of high-energy
observables, supersymmetry \emph{is} preserved.

Our original motivation for this work was to investigate
whether the well-known observation of supersymmetry for
asymptotic thermal masses,
in weakly coupled plasmas, did extend to other quantities.
We will answer this question in the affirmative, and will in fact
extend it to all high-energy correlators we could study.
More precisely, if high-energy
supersymmetry violations are characterized by the power
of the energy $E^{-n}$ by which they are suppressed
(relative to the vacuum correlators),
our finding is that $n> 2$ with strict
inequality in all considered cases.
This is a nonempty statement, since the leading thermal effects all
have $n\leq 2$.

We will discuss in turn the effective theories we have considered.
These include: the effective theory for particle masses at weak coupling,
in section \ref{sec:masses};
the effective theories for the imaginary part of self-energies
(including collinear bremsstrahlung processes and $2\to 2$ collisions),
at weak coupling, in section \ref{sec:imag}; 
the self-energies of uncharged particles in strongly interacting plasmas
with a gravity dual, in section \ref{sec:strong};
and finally, the operator product expansion for deeply virtual correlators,
in section \ref{sec:ope}.

We use the phrase ``effective theory'' to emphasize that
the details of the plasma are only probed through a restricted set
of low-energy correlators, which provide
parameters for medium-independent high-energy effective theories.
Supersymmetry should be understood as an intrinsic
property of these effective theories ---
we do not believe that
the thermal nature of the underlying medium plays any role.

\section{Thermal masses at weak coupling}
\label{sec:masses}

\begin{figure}
\begin{center}
\includegraphics[width=12cm]{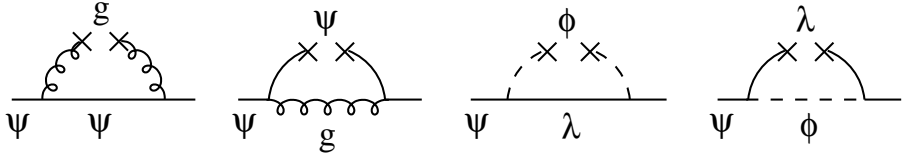}
\end{center}
\caption{One-loop fermion self-energy of a fermion $\psi$ due to gauge
interaction, at large energy $E$.
The plasma is probed by soft propagators
(with all components $\sim T$ in Minkowski spacetime),
which give rise to the low-energy correlators \refeq{masscondensates}
and are denoted here by the crosses.
The (hard) remainder of the diagram is expanded in powers of $T/E$.
}
\label{fig:masses}
\end{figure}

Thermal dispersion relations (of massless particles)
are known to approach
the form $E^2=p^2+m^2_{\infty}$ at large momenta $p\gg gT$
\cite{weldon}, at the leading order
in perturbation theory.
In applications to supersymmetric theories,
it has been repeatedly observed that the asymptotic masses $m_\infty$
are the same among particles within a supersymmetry multiplet%
\footnote{
Although this
observation has been described to me as ``well known''
in the course of several private conversations,
I did not succeed in finding a reference to it in the early
literature. A recent appearance is in \cite{susymass}.
}.
Compiling results from the literature \cite{pisarskimass} \cite{bosonicmasses},
or by directly evaluating one-loop diagrams such as those shown in
fig.~\ref{fig:masses},
one arrives at the formulae:
\bea
m^2_{\infty,g}=m^2_{\infty,\lambda}&=&
g^2C_A\left(Z_g + Z_f^\lambda\right)
+ g^2N_{\rm matter} T_M \left(Z_f^\psi + Z_S \right),
\nl
m^2_{\infty,\psi}=m^2_{\infty,\phi}&=& g^2C_M\left( Z_g + Z_f^\lambda +
Z_f^\psi + Z_S \right) + y^2\left( Z_f^\psi + Z_S \right),
\label{massmultiplet}
\eea
where the $Z_i$ are certain (non-local)
dimension-two condensates that we give shortly.
Terms in \eq{massmultiplet} are in one-to-one correspondence
with particles and interactions of renormalizable supersymmetric gauge
theories.
$C_{A}$, $C_{M}$ and $T_M$ are quadratic Casimirs and Dynkin indices
for the adjoint and matter representations, respectively,
and $N_{\rm matter}$ is the number of chiral superfields, with $g$ the
ordinary Yang-Mills coupling.
For simplicity the Yukawa contribution in \eq{massmultiplet}
is normalized to correspond to a term
$\sim \frac{y}{\sqrt2}\phi \,\psi\psi + {\rm c.c.}$ in the Lagrangian
of a single-field Wess-Zumino model.
We expect supersymmetry to be preserved for more general
(e.g. nonrenormalizable)
superpotentials, though we have not checked this explicitly.
Nonzero expectation values for the $D$ or $F$ auxiliary fields,
not considered in \eq{massmultiplet}, could break the supersymmetry
by lifting the bosonic masses%
\footnote{
Is is perfectly consistent to turn on a temperature without
generating expectation values for these fields.
For instance, at the leading order in perturbation theory,
$F\sim \phi^\dagger\phi^\dagger$ vanishes (it involves two
antiholomorphic fields),
and so does $D^a\sim \phi^\dagger t^a \phi$ for $t^a$ traceless.
}.
However, we do not view such effects as being specifically thermal
sources of supersymmetry violations, since they would do the same thing
in vacuum.

The condensates in \eq{massmultiplet}, each normalized to give the
contribution from two degrees of freedom, admit the following gauge-invariant
definitions and thermal expectation values:
\bea
Z_g \equiv \frac{1}{d_A}
\la v_\sigma F^{\sigma \mu} \frac{-1}{(v{\cdot}D)^2}
v_{\sigma'} F^{\sigma'}{}_\mu\ra
&=& 2\int \frac{d^3q}{(2\pi)^3} \frac{n_B(q)}{q} = \frac{T^2}{6},
\nl
Z_S \equiv \frac{2}{d_M} \la \phi^*\phi \ra &=& 2\int
\frac{d^3q}{(2\pi)^3}\frac{n_B(q)}{q}=\frac{T^2}{6},
\nl
Z_f^\psi \equiv \frac{1}{2d_M} \la \psibar \frac{\vslash}{v{\cdot}D} \psi\ra &=&
2\int \frac{d^3q}{(2\pi)^3}\frac{n_F(q)}{q}=\frac{T^2}{12},
\label{masscondensates}
\eea
with $v^\mu=(1,{\bf v})$ the four-velocity of the hard particle,
$n_{B,F}$ the standard Bose-Einstein and Fermi-Dirac distribution
functions, and $d_{A,M}$ the dimensions of the adjoint and matter
representations, respectively.
Useful examples include the thermal masses in ${\cal N}=4$ SYM,
which are all equal to $m_{\infty}^2=g^2N_c T^2$, and the gluon and
gluino masses
in pure glue SQCD, $m_{\infty,g(\lambda)}^2=\frac14 g^2N_c T^2$.

The structures in \eq{masscondensates} are identical
to those entering the hard thermal loop effective action
\cite{braatengen,taylorwonggen}:
these are in fact the unique dimension-two gauge-invariant
operators that can be built out of a light-like four-vector $v^\mu$
\cite{braatengen}.
Given that next-to-leading order ($\OO(g)$) corrections
are associated with hard thermal loop physics \cite{htlpaper}
($gT$ scale physics), but only $\sim g^2$ corrections
arise from the hard scale $\sim E$,
one concludes that at $\OO(g)$ only the matrix elements
\refeq{masscondensates} get corrected but not
their coefficients \refeq{massmultiplet}.
Thus, $\OO(g)$ corrections preserve supersymmetry.
For completeness, we record their values here.  They
are most readily obtained by working within the
dimensionally reduced Euclidean theory \cite{dimred},
in which the corrections to $Z_g$
only arise from the contribution $Z_g\to \frac{1}{d_A} 2A_4^a A_4^a$
of longitudinal gluons with zero Matsubara frequency.
Next-to-leading order thermal masses were also
obtained in \cite{blaizot00} by means of an indirect
thermodynamic argument, by relating them
to the well-known $\sim g^3T^3$ corrections to the QCD entropy%
\footnote{
The authors of \cite{blaizot00} also expressed worries that
actual thermal dispersion relations of hard particles could grow like
$m^2\sim ET$ at high energies, which would spoil the physical
interpretation of the ``effective'' thermodynamic masses they have computed.
This was motivated by a partial evaluation of the gluon self-energy
which included only its longitudinal contribution.
I believe that such a growth is not realized
in a complete calculation.
Indeed, in the effective theory language we are using, the worrisome
contribution (A17) in \cite{blaizot00} can be matched to a matrix
element of the dimension-1 gauge-invariant gluon condensate
$i v{\cdot}A \delta(-iv{\cdot}D) v{\cdot} A$, which arises from the
eikonal coupling of charged particles to gauge fields.
Although this operator definitely contributes to the imaginary
part of self-energies --- it gives rise to the universal
$\Gamma\sim g^2T\log(1/g)$ damping rate of charged particles
\cite{pisarskimfp} --- as I will discuss more fully
in a forthcoming paper \cite{toappear}, the longitudinal
and transverse gluon contributions exactly cancel out
from its real part.
This cancellation can be foreseen through a causality argument:
the real part (of a time-ordered correlator, which is the natural
operator ordering at high energies)
is equal to that of a retarded correlator, which cannot probe
the plasma at light-like separations.
Thus, the thermodynamic results of \cite{blaizot00}
ought to agree with \eq{massnlo}.
\label{foot:dim1}
}.
Either way one obtains ($Z_f^{\rm NLO}=Z_f^{\rm LO}+\OO(g^2T^2)$):
\be
Z_g^{\rm NLO} = \frac{T^2}{6}-\frac{Tm_{\infty,g}}{\pi\sqrt{2}} + \OO(g^2T^2)\,,\hspace{1cm} 
Z_S^{\rm NLO} = \frac{T^2}{6}-\frac{Tm_{\infty,S}}{2\pi} + \OO(g^2T^2)\,.
\label{massnlo}
\ee

What the
factorization formula \refeq{massmultiplet} (and the fact that it
preserves supersymmetry)
becomes when $\OO(g^2)$ effects
are accounted for, if it retains any meaning at all, is unknown.
The derivation outlined in fig.~\ref{fig:masses} shows that the
supersymmetry of the thermal masses at leading order may be
interpreted as a statement about the
couplings of soft particles with various spins to hard
propagators: these turn out to be universal in supersymmetric theories.


\section{Imaginary parts of self-energies at weak coupling}
\label{sec:imag}

The imaginary parts of self-energies at weak coupling arise from
$2\to 2$ scattering against plasma particles, as well as from induced
collinear splitting processes (bremsstrahlung or pair production).
For charged particles in gauge theories,
the dominant contribution to $\Im \Pi$ is $\sim g^2TE$ due to
elastic small angle Coulomb scattering, though the
dominant inelastic contribution $\sim g^4T^{3/2}E^{1/2}$ (barring
logarithms) is due to induced collinear processes, which we will discuss
first.  These processes also dominate the self-energies of neutral
particles in gauge theories, provided these particles are allowed
to split into charged ones.
In non-gauge theories, self-energies begin at $\sim g^4T^2E^0$
due to ordinary $2\to 2$ scattering, which we will discuss in subsection
\ref{sec:2to2}.

\subsection{Collinear pair production}
\label{sec:rates}

The key aspects of collinear pair production may be briefly
summarized as follows.
This process is only relevant in gauge theories, in which it is initiated
by the very frequent small-angle (Coulomb) scatterings suffered by either the
parent or the daughter particles.
At high-energies $E\gg gT$, its long formation time
(associated with its collinearity) requires that
multiple thermal scatterings
occurring during it be summed coherently.  This
causes a parametrically significant destructive interference,
the so-called LPM effect \cite{LPM}, which is responsible for the
non-analytic behavior $\Pi \propto E^{1/2}$.
For relativistic plasmas, a complete
leading-order treatment was given (for photons) in \cite{AMYphoton} (see also
earlier discussions \cite{BDMPS,Zakharov}, in which different
approximations are made).
Somewhat schematically, the result may be written in the form:
\be
-2{\rm Im}\, \Pi_a(E) = \sum_{bc} \int_0^1 dz P_{a\to bc}(z) F_{a\to bc}(E,z)
\label{susyrate}
\ee
where $P_{a\to bc}$ are ordinary DGLAP kernels \cite{DGLAP},
governing collinear physics, $bc$ indexes final states
and $z=E_b/E_a$ is the longitudinal momentum fraction.
In \eq{susyrate} we have omitted final state Bose-enhancement or
Pauli-blocking factors which are not needed unless $z$ or $(1-z)$
are very small, $\sim T/E$.
The functions $F(E,z)$ depend in a complicated way on $E$ and $z$
and are to be obtained by solving
an effective inhomogeneous Schr\"odinger
equation governing the evolution of the pair
in the transverse plane \cite{AMYphoton}.
This equation depends on the details of the plasma through a collision
kernel $C(q_\perp)$, which is a function of the transverse momentum transfer.
Its only property that we need
is that it involves only eikonal
physics: it does not depend on the spins of the
particles.
For our purposes $F(E,z)$ in \eq{susyrate} is thus just some universal
function, the same for all final states among a given
supersymmetry multiplet.
In the leading logarithmic approximation \cite{AMYphoton},
$F(E,z)\sim g^4N_c^2T^{\frac32}E^{\frac12}
z^{-\frac12}(1-z)^{-\frac12}
(\log(\frac{ET}{g^2T^2z(1-z)}))^{1/2}$.

The only ingredients in \eq{susyrate} which could
possibly break supersymmetry are the DGLAP splitting
kernels $P_{a\to bc}(z)$.
Such kernels are listed in table \ref{tab:DGLAP}, for various
supermultiplets of initial and final states.
As shown in the table, when complete supermultiplets of
final states are summed over (thereby enforcing the
symmetry under $z\to (1-z)$),
supersymmetry with respect to the initial particle is restored.
Not shown in the table (it is related to the first three entries by a
crossing
symmetry \cite{dokshitzer}), but which also preserves supersymmetry,
is the process of bremsstrahlung of a gauge particle off a chiral multiplet.
Thus, all in-medium splitting rates preserve supersymmetry.

Observations of supersymmetry in DGLAP kernels
were made long ago by Dokshitzer \cite{dokshitzer}, and
subsequently given an explanation (at the one-loop level, we believe)
by Lipatov and collaborators
\cite{lipatovearly}.
Here we are merely reporting their implications
at finite temperature.

\begin{table}
\begin{center}
\begin{tabular}{|c|c|c|} \hline
Process & DGLAP kernel $P(z)$ & Sum \\ \hline
$\gamma\to \psi^\dagger\psi$ & $e^2\left[z^2+(1-z)^2\right]$
& $e^2$
\\
$\gamma\to \phi^\dagger\phi$ & $e^2\left[2z(1-z)\right]$
& \\ \cline{2-3}
$\tilde{\gamma}\to \phi^\dagger \psi$ & $e^2\left[2z\right]$
& $e^2$
\\
\hline
$g\to gg$ & $g^2C_A\left[ 2(1-z)/z + 2z/(1-z) +2z(1-z)\right]$
& $g^2C_A\left[2/z+2/(1-z)-3\right]$
\\
$g\to \lambda^\dagger\lambda$ & $g^2C_A \left[z^2+(1-z)^2\right]$
& \\ \cline{2-3}
$\lambda\to g\lambda$ & $g^2C_A\left[4z/(1-z) + 2(1-z)\right]$
& $g^2C_A\left[2/z+2/(1-z)-3\right]$
\\
\hline
$\phi\to \psi^\dagger \psi^\dagger$ & $y^2\left[1\right]$
& $y^2$
\\ \cline{2-3}
$\psi\to \phi^\dagger\psi^\dagger$ & $y^2\left[2z\right]$
& $y^2$
\\
\hline
\end{tabular}
\end{center}
\caption{DGLAP splitting kernels.
The first three entries govern the splitting of photons and photinos
to a chiral multiplet,
the next three pertain to the Yang-Mills splitting of a
nonabelian gauge multiplet,
and the last two govern the
Yukawa splitting of a chiral multiplet.
Supersymmetry is restored
when complete supermultiplets of final states are summed over.
}
\label{tab:DGLAP}
\end{table}

We expect coupling constant corrections to \eq{susyrate}
to first arise at $\OO(g)$.  In thermal perturbation theory
such $\sim g$ factors arise from
ordinary loop factors $g^2$ multiplied by large bosonic
occupation numbers $n_B\sim T/p^0\sim T/gT$, and are associated with
hard hard thermal loop physics \cite{htlpaper} at
the $gT$ scale.
Such soft physics can only interfere with processes having a sufficiently
long duration, such as the soft scatterings contributing to $C(q_\perp)$
with $q_\perp\sim gT$, so we believe that this is the only ingredient
suffering from $\OO(g)$ corrections.
These soft collisions have a purely diffusive effect, so we could
equivalently say that all $\OO(g)$ corrections pertain to the so-called
transverse momentum diffusion coefficient ``$\hat{q}$''.
These corrections should be calculable using techniques
similar to those used for heavy quark diffusion in \cite{nlodiffusion},
but this has not yet been done.
Since only eikonal physics is involved, these corrections
trivially preserve supersymmetry.

The $\OO(g^2)$ corrections to \eq{susyrate} are expected to possess
a much more interesting and richer structure.
For instance, they will most certainly require dealing with the scale
dependence
of the partonic constituents of the plasma, which should
ultimately lead to ``saturation'' effects \cite{saturation}
at very high energies upon
summation of large logarithms $\alpha_s\log(E/T)$
and $\alpha_s\log(q_\perp^2/T^2)$ with $q_\perp^2\sim E^{1/2}T^{3/2}$.
The scale evolution of the
constituents of the probe, which has to be treated
in the presence of the LPM effect,
should also enter at this order.
Other interesting (though manifestly supersymmetry-preserving) effects
may include sensitivity to nonperturbative $g^2T$-scale magnetic
physics, which we believe contributes to $\hat{q}$ at $\OO(g^2)$.
We leave to future work a detailed analysis of these effects
and of the question of whether they preserve supersymmetry.

As for the subleading corrections in $T/E$,
we expect supersymmetry-breaking effects in $\Pi$ not to be
larger than $\sim T^{5/2}E^{-1/2}$.  These could arise
from various $\sim T/E$ or $\sim q_\perp^2/E^2\sim (T/E)^{3/2}$
corrections to ingredients entering $F(E,z)$,
such as the eikonal vertices.

\subsection{$2\to 2$ scattering at weak coupling}
\label{sec:2to2}

\begin{figure}
\begin{center}
\includegraphics[width=10cm,height=3cm]{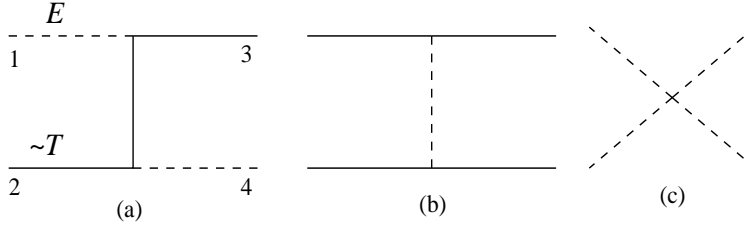}
\end{center}
\caption{$2\to 2$ scattering processes in Wess-Zumino model; solid lines
  are fermions and dashed lines are scalars.
}
\label{fig:wz}
\end{figure}

Ordinary $2\to 2$ collisions dominate self-energies in non-gauge models,
which we will now discuss; their total rate is found to preserve
supersymmetry.
We first recall the general formula for the total collision rate
($-\Im \Pi=-\Gamma E$):
\bea -2\Im \Pi(p_1) &=&
\int \frac{d^3p_2d^3p_3d^3p_4}{(2\pi)^{12} 2E_2 2E_3 2E_4}
(2\pi)^4\delta^4(p_1+p_2-p_3-p_4) \\
&& \hspace{0.5cm} \times
\sum_{s_{2}s_{3}s_{4}} \left|{\cal M}_{1s_{2}\to s_{3}s_{4}}\right|^2 n_b(E_2)(1\pm n_c(E_3))(1\pm
n_d(E_4)).
\label{genericrate}
\eea
Here the particle labels are as defined as in
fig.~\ref{fig:wz}, the $s_{i}$ label the corresponding particle species,
and $n_{i}$ are the corresponding distribution functions.

Let us first assume, for a moment, that the distribution functions can be
omitted in the final state (``Bose-enhancement'' and ``Pauli-blocking'')
factors
$(1\pm n_i)$, which is justified for generic final state energies
$E_3\sim E_4\sim \sqrt{E_1 E_2}\sim \sqrt{ET}$.
The integrand then depends only on the sum $\sum_{s_{3}s_{4}} |{\cal
  M}|^2_{1s_{2}\to s_{3}s_{4}}$.
Such matrix elements summed over final states turn out to
obey supersymmetry
identities, with respect to the particle $1$, for fixed identities of
particle $2$. This is exemplified
in table~\ref{tab:matrix} for
single-field Wess-Zumino model with cubic superpotential,
and the generalization to other models will be given shortly.
Therefore, the contribution to \eq{genericrate} from the region $E_3,E_4\gg T$
preserves supersymmetry.

\begin{table}
\begin{center}
\begin{tabular}{|c|c||c|c|} \hline
Process & $|{\cal M}|^2/4y^2$ & Processes & $|\overline{{\cal
    M}}|^2/4y^2$ \\ \hline
$\psi\psi\to\psi\psi$ & $1$
& $\psi \psi\to X$, $\phi\psi\to X$ & $1$
\\
$\phi\phi\to\phi\phi$ & $1$
& $\psi \psibar\to X$, $\phi\psibar\to X$ & $\left[2+\frac{u}{t}+\frac{t}{u}\right]$
\\
$\phi\psi\to\phi\psi$ & $-u/s$
& $\psi \phi\to X$, $\phi\phi\to X$ & $1$
\\
&
& $\psi \overline{\phi}\to X$, $\phi \overline{\phi}\to X$ & $\left[2+\frac{u}{t}+\frac{t}{u}\right]$
\\ \hline
\end{tabular}
\end{center}
\caption{Left panel: scattering amplitudes $|{\cal M}|^2$ in
  Wess-Zumino model, with amplitudes related by crossing symmetry not shown.
Right panel: amplitudes summed over final states, for which
supersymmetry is manifest as a function of particle $1$ with particle
  $2$ held fixed.
}
\label{tab:matrix}
\end{table}

It is easy to convince oneself that for bounded amplitudes
$|{\cal M}|^2$,
the regions $E_3\sim T$ and $E_4\sim T$ suffer from $\sim T/E$
phase-space suppressions, and so the neglect of the final state
distributions in \eq{genericrate} is justified.
However, $s/t\sim ET/T^2$ singularities in squared matrix
elements when $t\lsim T^2$ can overcome
this suppression, and a separate discussion is required
for these singular terms%
\footnote{
The total integral of such $\sim 1/t$ singularities
is logarithmically divergent at $t\to 0$.  This
is cured by resumming hard thermal loop self-energies \cite{htlpaper}
to the soft exchanged fermion propagator.
}
(The physically equivalent region $u\to 0$, or $E_3\sim T$, can be
treated similarly).
To establish the supersymmetry of this region, for which the
distribution function $n(E_4)$ must be kept, we need
another ingredient: universality of the $1/t$ singularities.
Indeed, the coefficient of $1/t$ at $t\to 0$, which is due to soft
fermion exchange, is left unchanged when the hard
particle $1$ is replaced by its
superpartner (e.g. if particles $1$ and $3$ are exchanged in
fig.~\ref{fig:wz}).
This shows that the complete $\sim T^2E^0$ self-energies in the Wess-Zumino
model preserve supersymmetry, up to $\sim T^3E^{-1}$ corrections.

It is interesting to analyze the $t\to 0$ singularities by means
of the effective theory language used for thermal masses in
section \ref{sec:masses}.
Indeed, the region $E_4\sim T$, $t\sim T^2$
in fig.~\ref{fig:wz} is characterized by soft fields coupled to
a hard line and is thus governed by the gradient expansion
depicted in fig.~\ref{fig:masses}.
This means that the $\sim T^2E^0$ contribution to \eq{genericrate}
from soft fermion
exchange may be understood as a contribution to the imaginary part
of the fermion condensate in \eq{masscondensates}, at one-loop
in thermal perturbation theory%
\footnote{
For scalar exchange, this argument relates the locality of the scalar
condensate $\phi^*\phi$ in \eq{masscondensates} (which implies
that it is purely real) to
the absence of $1/t$ singularities due to scalar exchange.
}.
This way we see that the
universality of $1/t$ singularities as a function
of particle $1$ is closely linked to the supersymmetry of
thermal masses.

We now show, as promised, that supersymmetry of scattering
amplitudes summed over final states holds in any supersymmetric theory,
as a property of the S-matrix.
Introducing the notation $P_{i_1\ldots i_n}$ for projection operators
which perform the sum over complete supermultiplets of
scattering states with $n$ particles,
this follows from considering the following trace (over scattering
states):
\be \Tr \left[ S^\dagger P_{34}S  \left(|2\rangle\langle 2|
\otimes \left[Q,|1\rangle\langle
    \tilde{1}|\right]\right) \right],
\label{susysum}
\ee
with $S$ the S-matrix and $\tilde{1}$ denotes the superpartner of particle $1$.
For any supersymmetry generator $Q$ which does not annihilate particle $1$,
the commutator $[Q,|1\rangle\langle \tilde{1}|]\propto
(|1\rangle\langle1|-|\tilde{1}\rangle\langle{\tilde{1}}|)$,
so \eq{susysum} computes the difference:
\be \sum_{s_3,s_4} \left(
\left|M_{12\to s_{3} s_{4}}\right|^2
-\left|M_{\tilde{1}2\to s_{3} s_{4}}\right|^2\right). \label{susysum2}
\ee
For a massless particle $2$ one can always choose $Q$
so as to annihilate particle $2$; such a $Q$ commutes
with $|2\rangle\langle 2|$, with the S-matrix as well as with
the projectors $P_{i_1\ldots i_n}$ (by construction), showing
that \eq{susysum} (and thus
\eq{susysum2}) vanishes, being the trace of a commutator.
This establishes the supersymmetry of the contributions from
$E_3,E_4\gg T$ to \eq{genericrate} in any theory%
.

Combining the results of the preceding sections, we have reached
a simple conclusion:
the full thermal self-energies of gauge-neutral particles
preserve supersymmetry, at leading order in the coupling,
up to corrections suppressed by a at least $T^{5/2}E^{-1/2}$.
Although we believe the analysis can be generalized to
charged particles (for which the analysis is made more
complicated by the stronger singularities ${\cal M} \sim 1/t$
associated with gluon exchange%
\footnote{
In an effective theory language this implies the existence of certain
pure imaginary operators
which are not probed by the thermal masses,
such as the dimension-1 operator described in footnote{foot:dim1}.
At dimension-2 I find
operators like $v{\cdot}A\delta(-iv{cdot}D) D$
(representing e.g. an interference term
between t-channel gluon exchange and $D$-term
scalar self-interaction in $\phi\phi\to \phi\phi$ scattering), as well
as its superpartner involving $\lambda$.
At least for vanishing chemical potential I could check that
such operators take on no expectation value.
},
and various sources of infrared divergences which make these self-energies
less cleanly defined),
here we will refrain from doing so: we are
content with a robust result for gauge-invariant self-energies.

\section{Strong coupling}
\label{sec:strong}

Maldacena's conjectured gauge/gravity correspondence \cite{adscft} renders possible,
among other things,
the calculation of
correlators of currents in certain strongly coupled
large $N_c$ gauge theories.
In theories which have a continuous R-symmetry, such as the SU(4)
of ${\cal N}=4$ super Yang-Mills, ``photons'' and
``photinos'' can be introduced by weakly gauging a U(1) subgroup of the
R-symmetry.
Their self-energies are then given by suitable two-point
functions of currents and their superpartners,
which we now evaluate by means of the correspondence.

In the case of the on-shell photon self-energy in ${\cal N}=4$ SYM,
it was argued by means of a WKB approximation \cite{kovtunstarinetsetal}
(in appendix)
that at high energy
the calculation localizes itself near the boundary of the AdS space.
Here we generalize this phenomenon to other backgrounds,
and establish two properties of the resulting effective theory of
high-energy photon/photino propagation.
First, it only probes the underlying low-energy medium through the
expectation value of its energy-momentum tensor
(actually, only through one component $\propto p_\mu p_\nu T^{\mu\nu}$).
Second, it preserves supersymmetry:
the absorption rates and dispersion relations
of a photon and of a photino are identical.

We will be considering five-dimensional metrics of the general form
\be
ds^2 = R^2\frac{g(z)dz^2 + h_{\mu\nu}(z)dx^\mu dx^\nu}{z^2}\,,
\label{metric}
\ee
for which, near the boundary $z=0$, the metric approaches
that of ${\rm AdS}_5$ with radius $R$ (for which $g(z)= 1$ and
$h_{\mu\nu}(z)= \eta_{\mu\nu}$).
The metric \eq{metric} should be sufficiently general to cover any system
invariant under space-time translation that admits a gravity dual.
For the ${\rm AdS}_5$ black hole, relevant for ${\cal N}=4$ SYM at
finite temperature $T$, $-h_{00}=1-(\pi T z)^4$, $h_{ij}=\delta_{ij}$,
$h_{i0}=0$ and $g(z)=(-h_{00})^{-1}$.
At certain steps below rotational invariance will be assumed;
these steps will be highlighted.

\subsection{Bulk equations}

The bulk dual of the spin-1 current which couples to the photon
is a five-dimensional
gauge field, whose field strength tensor obeys Maxwell's equations:
\begin{align}
0&=\frac{z}{\sqrt{g(z)\det(-h(z))}}
\partial_z \left( \frac{h^{\nu\sigma}}{z} \sqrt{\frac{\det(-h(z))}{g(z)}} F_{z\sigma}\right)
+ h^{\nu\sigma} h^{\mu\rho} \partial_\mu F_{\rho\sigma}\,,
\\
\partial_\alpha F_{\mu\nu} &= \partial_\mu F_{\alpha\nu}-
\partial_\nu F_{\alpha\mu}\,.
\end{align}
Here $\mu,\nu,\sigma,\rho$ are space-time indices
but $\alpha$ may cover all five coordinates.
We will restrict our attention to space-time
momentum eigenstates $\partial_\mu=ip_\mu$.
A closed equation for the transverse electric field $F_{0\perp}$,
for $\nu=\perp$ a component perpendicular to $p_\mu$,
may be obtained by acting on the first equation
with a partial time derivative $\partial_0$, and using the second
equation.  Specifically, one uses relations such as
$\partial_0 F_{z\perp}=\partial_z F_{0\perp}$, which follow from
dropping perpendicular derivatives $\partial_\perp$ in the latter.
To fully exploit such simplifications,
rotational invariance must be assumed, so that upstairs derivative
$h^{\perp\sigma}\partial_\sigma$ also vanish.
This yields the closed equation:
\be
\frac{z h_{\perp\perp}}{\sqrt{g(z) \det(-h(z))}} \partial_z
\left( \frac{h^{\perp\perp}}{z}
\sqrt{\frac{\det(-h(z))}{g(z)}}
\partial_z F_{0\perp}\right)
= h^{\mu\nu}p_\mu p_\nu
F_{0\perp}, \label{maxwelleq}
\ee
in which no summation over $\perp$-indices is implied.

The bulk dual of the spin-$\frac12$ operator
coupling to the photino
is a five-dimensional Dirac fermion with mass $m=\frac12$
\cite{spinorads} (in units with $R=1$).
It possesses as many components as two four-dimensional Weyl spinors,
but it is dual to only one such spinor: the sign of $m$ breaks
the symmetry between the two Weyl components.
The bulk Dirac equation reads:
\be \left[\Dslash + m\right]\psi = 0 \equiv
\left[ \gamma^a e_a^\alpha \left(\partial_\alpha +\frac14
  \omega_\alpha {}^{ab} \gamma_a \gamma_b\right) + m\right] \psi\,,
\ee
with $\alpha,a=0\ldots4$ and $e_a^\alpha$ the orthogonal basis.
Under the assumption of rotational invariance,
the term involving the spin connection $\omega$
must be proportional to the single matrix $\gamma_z$, and it
can be removed by a $z$-dependent field rescaling.
We choose the rescaling
$\psi= z^2(\det (-h))^{-1/4}e^{-m\int^z dz \, \sqrt{g(z)}/z}\tilde{\psi}$,
which leads to the following equations for the Weyl components of
$\psi_{L,R}$ of $\tilde{\psi}$:
\bea
\partial_z \psi_L = \sqrt{g(z)}\pslash_R \psi_R,
\label{dummydirac0} \\
\left[ \frac{1}{\sqrt{g(z)}} \partial_z - \frac{2m}{z}\right] \psi_R
= \pslash_L \psi_L.
\label{dummydirac1}
\eea
Here $\pslash_{L,R}$ are Weyl operators associated with
the four-dimensional metric $h_{\mu\nu}(z)$.
With $m=+\frac12$ the component relevant near the $z=0$ boundary
is $\psi_L$ and we are calculating the self-energy of a left-handed
photino.
\Eq{dummydirac1} implies a closed equation for $\psi_L$:
\be
\pslash_R \left[ \frac{1}{\sqrt{g(z)}}\partial_z
 - \frac{2m}{z}\right]\frac{1}{\pslash_R \sqrt{g(z)}}
\partial_z \psi_L
= h^{\mu\nu}p_\mu p_\nu \psi_L \,. \label{diraceq}
\ee

\subsection{WKB solution and supersymmetry}

We are now in position to discuss the WKB approximation.
By means of a change of variable $y\equiv y(z)$,
Maxwell's equation \eq{maxwelleq} may be cast in a Schr\"odinger
form with potential proportional to the squared energy $p_0^2$,
provided
\be
\frac{d y}{dz} = 2zh_{\perp\perp} \sqrt{\frac{g(z)}{\det (-h(z))}} \,.
\label{changevar}
\ee
The factor of $2$ is convenient:
near the boundary $y\sim z^2$.
For black holes (like the ${\rm AdS}_5$ black hole metric given
above) the function $g(z)$ has a pole at a
finite value of $z$ (the location of the horizon), while the function
$\det (-h)$ vanishes there.  In this limit
$y$ is mapped logarithmically to infinity.  The rescaled potential
remains finite there, though, due to the blowing up of $h^{00}$;
it then depends only on the energy $E=p^0$.

\begin{figure}
\begin{center}
\includegraphics[width=5.5cm]{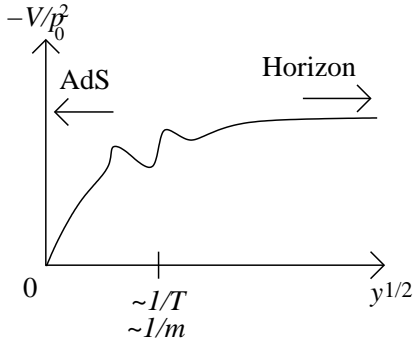}
\end{center}
\caption{Schematic features of the
Schr\"odinger potential $V(y)/p_0^2$, when $p^2=0$.
It approaches the universal linear behavior \eq{bdypot} near the boundary
and tends to a constant at the horizon $y\to \infty$, with
a transition regime that may depend on the details of the theory
and on possible intrinsic mass scales $m$.
}
\label{fig:pot}
\end{figure}

The qualitative features of the Schr\"odinger potential in
the equation
$[\partial_y^2 - V(y)]F_{0\perp}=0$ are sketched in fig.~\ref{fig:pot}.
The shape of the potential depends on the geometry but not on the
energy, which only determines its overall normalization.
At large $y$ the potential becomes constant,
while for $y\to 0$ the leading term becomes,
for on-shell and off-shell momenta respectively,
\be
V(y)\to\left\{
\begin{array}{ll} \displaystyle
\frac{1}{4y} p^2, & p^2\neq 0, \\ \displaystyle
\frac{y}{4} p_\mu p_\nu \frac{d \,h^{\mu\nu}(z)}{d(z^4)} = -y \frac{
  \pi^2T^{\mu\nu}p_\mu p_\nu}{2N_c^2}, & p^2=0. \label{bdypot}
\end{array}\right.
\ee
Here we have used that the leading corrections to the
metric near the boundary are proportional to $z^4$ and are
related to the expectation value of the stress-energy tensor $T_{\mu\nu}$;
its trace part, if nonzero, does not contribute when $p^2=0$.
The normalization in \eq{bdypot} is appropriate to the ${\cal N}=4$ SYM theory.

At the horizon $y\to\infty$ the solutions are oscillatory
and in-falling boundary conditions
$F_{0\perp}\propto e^{i\omega}$ must be imposed for calculating
retarded correlators \cite{sonstarinets}, with $\omega= p^0\pi T/2$
the natural frequency near the horizon.
To obtain correlators of currents, as described shortly,
this solution must be continued to the ${\rm AdS}_5$ boundary $z=0$.
For energies high relative to all intrinsic scales in the metric,
a WKB approximation can be used.  This is applicable
for $y$ down to $y \sim 1/p^2$ (resp. $y \sim (T^4E^2)^{-1/3}$)
for $p^2\neq0$ (resp. $p^2=0$), at which it breaks down
due to the redshift factors%
\footnote{
The transition between the two regimes $p^2\neq0$ and $p^2=0$
takes place smoothly around $|p_s^2|\sim E^{2/3}T^{4/3}$,
for which value of $p^2$ the two estimates for $y$ cross each other.
This natural scale $p_s^2(E)$ is exactly the
``saturation scale'' $p_s\sim T/x_s$
discussed in \cite{iancuetal} at strong coupling,
viewed as a function of $E$ with $x_s\equiv p_s^2/2ET$.
}.
Note that these scales are the intrinsic scales of the Schr\"odinger
equations with approximate potentials \refeq{bdypot}.
The problem is thus reduced to solving
those approximate equations, with large $y$ behavior matching the WKB
form $\propto V^{-1/4}e^{i\int^y dy \sqrt{V}}$.

For the Dirac equation \eq{diraceq} one needs only modify
the change of variable \eq{changevar}, to:
\be \frac{dy}{dz} = 2z \sqrt{g(z)} \frac{\pslash_R(z)}{\pslash_R(z=0)}.
\label{fermvar}
\ee
For the on-shell component $\psi_L^-$ of a left-handed photino in a
rotationally-invariant background, the operator
$\pslash_R$ is nonsingular with eigenvalue
$E(\sqrt{|h^{00}|}+\sqrt{h^{33}})$.  Here $h^{33}$ gives the metric
component along the longitudinal direction.
Just as for \eq{changevar}, near the boundary $y\sim z^2$
and the horizon is mapped logarithmically to $y=\infty$,
so the same WKB approximation applies.
More significantly, one readily sees from \eq{diraceq}
that the approximate potentials near the boundary are
\emph{identical} to the photon case, \eqs{bdypot}.

Correlation functions are obtained by prescribing the limiting
values of the fields $F_{0\perp}$
and $\psi_L$ near the boundary and evaluating boundary terms
$\propto \partial_y F_{0\perp}$ (see e.g. \cite{kovtunstarinetsetal}),
or proportional to $\psibar\psi\sim \psi_R/z \sim
\frac{1}{\pslash_R}\partial_y \psi_L$ \cite{spinorads},
or more precisely,
\begin{align}
\Pi_{\gamma}=
-\frac{N_c^2T^2}{8\pi^2}\lim_{y\to 0}\frac{\partial_y
  F_{0\perp}(y)}{F_{0\perp}(y)},
&\hspace{0.5cm}
\Pi_{\tilde{\gamma}}=
-\frac{N_c^2T^2}{8\pi^2}\lim_{y\to 0}\frac{\partial_y
  \psi_L^-(y)}{\psi_L^-(y)}.
\label{bdyterm}
\end{align}
Here $\Pi_{\tilde{\gamma}}\equiv \overline{u} \Sigma u$ is the photino self-energy
sandwiched between on-shell polarization spinors $u$, whose real part
yields the thermal mass squared.
The normalizations \eq{bdyterm} are fixed
by the well-known supersymmetry-preserving vacuum result,
$\Pi_\gamma=\Pi_{\tilde{\gamma}}=-N_c^2p^2/32\pi^2 \log(p^2/\mu^2)$,
$p^2={\bf p}^2-p_0^2$.
On the light-cone, Schr\"odinger's equation with the approximate
potential \refeq{bdypot} is solved in terms of Bessel (Hankel) functions,
\be
F_{0\perp}(y)=\psi_L^- =
y^{\frac12}\left(J_{\frac13}\left(\frac23 \omega y^{\frac32}\right)
+i Y_{\frac13}\left(\frac23 \omega y^{\frac32}\right)\right),
\ee
with $\omega^2= \pi^2T_{\mu\nu}p^\mu p^\nu/2N_c^2$, yielding with
\eq{bdyterm} the result:
\be
\Pi_\gamma(p)=\Pi_{\tilde{\gamma}}(p) = \frac{N_c^2
\Gamma\left(\frac23\right)}{16\pi^2\Gamma\left(\frac13\right)}
\left(3^{\frac13}-i3^{\frac56}\right) \omega^{\frac23}
,\hspace{1cm} \mbox{$p^0=p$ large.} \label{strongres}
\ee
The imaginary part of this result reproduces that given in
\cite{kovtunstarinetsetal}, with $\omega=p^0T^2\pi^2/2$ in
thermal ${\cal N}=4$ SYM.  Corrections in $T/E$
to this result may be found by expanding the potential
\eq{bdypot} to higher orders near the boundary; for the ${\rm AdS_5}$
black hole this expansion proceeds in powers of $y^2\sim \omega^{-4/3}$,
so the first subleading corrections to $\Pi$ are $\sim \omega^{-2/3}$.

It is remarkable that photon self-energies at strong coupling,
and high energies, depend only on \emph{one} property of the
plasma: its stress-energy tensor.
An heuristic picture of strongly coupled plasmas,
based on the idea of parton saturation,
has been proposed recently \cite{iancuetal}
in which such a universality comes out naturally.

\section{Operator Product Expansion}
\label{sec:ope}


Supersymmetry violations in deeply virtual correlators
may be analyzed by means of the operator
product expansion (OPE) \cite{ope}. 
The OPE is a means of separating short-distance and long-distance
physics, so that the thermal corrections to
deeply virtual (short-distance)
correlators with $E\gg T$
are expressed in terms of the expectation value of local operators.
Thermal corrections are thus suppressed by powers $\sim E^{-\Delta}$, with
the powers $\Delta$ set by the scaling dimensions of local operators%
\footnote{
It is worth noting that the OPE, being most rigorously formulated in Euclidean
signature, does not enjoy at the moment the same rigorous status
in Minkowski space-time with respect to nonperturbative physics
(see e.g. \cite{shifman00}).
}.

The difference between a correlator
of operators and of their superpartners
may be written as a supersymmetry variation
(which corresponds to the fact that
it vanishes in supersymmetry-preserving vacua).
For instance, for correlators of transverse
currents $\epsilon{\cdot}J$ and of their superpartners
$\lambda_\alpha$ \cite{sohnius}, one schematically has:
\be
\epsilon_1{\cdot}J(p) \,\epsilon_2{\cdot}J - \frac12
\lambda^\dagger(p) \,/\!\!\!\epsilon_1 \pslash \,/\!\!\!\epsilon_2 \lambda
\propto \epsilon_1^{\alpha\dot{\alpha}} Q_{\alpha}\left[
\lambda^\dagger_{\dot{\alpha}}(p) \epsilon_2{\cdot}\right]
\label{susyoff}\,
\ee
with $p_\mu\epsilon^\mu_{1,2}=0$
and $\alpha,\dot{\alpha}$ spinor indices.
As an operator equation the OPE must commute with the supersymmetries,
so from the OPE of the right-hand side of \eq{susyoff}
one concludes that the operators on its left-hand side are
\emph{supersymmetry variations}.
Thus, in order to see supersymmetry violations at order $E^{-2}$
or stronger, one
must find \emph{local} gauge-invariant fermionic operators of
dimension $\frac32$ or less.

In a wide class of theories, an accidental symmetry comes into play:
such operator do not exist.
These theories certainly include all weakly coupled
gauge theories containing no U(1) vector multiplets
and no gauge-neutral chiral superfields.
In these theories, gauge-invariant dimension-2 bosonic operators
(such as $\Tr \phi\phi$ or $\Tr \phi^*\phi$)
do not correspond to any supersymmetry variations, and must thus
enter in the same way in the OPEs of fields and of their
superpartners; this is easily verified at leading order
for the OPE of currents in weakly coupled gauge theories.
The lowest-dimensional
fermionic operators are dimension-$\frac52$ supercurrents,
from which we conclude that
thermal supersymmetry breaking can only be seen through dimension-3
operators, $\sim E^{-3}$.

When neutral chiral superfields or U(1) vector
multiplets are present, nonzero expectation values for
$D\sim \phi^*\phi$ or $F\sim \phi^*\phi^*$ auxiliary fields
(which enter the supersymmetry transformations of the
gaugino and fermionic matter fields, respectively)
could produce supersymmetry violations at dimension 2.
A similar possibility was observed for thermal masses,
in section \eq{sec:masses} but, as we discussed there,
we do not view it as being specifically related to thermal effects.
Thus, we conclude that in weakly coupled theories, the absence
of thermal supersymmetry breaking below dimension 3 is generic.

It is not possible to analyze general theories at finite
values of the coupling constants, because finite
anomalous dimensions can alter the power-counting.
Nevertheless, for certain strongly coupled
theories accessible to the AdS/CFT
correspondence, it is easy to be more quantitative.
For instance, it is known \cite{aharonyetal}
that in ${\cal N}=4$ SYM at large 't Hooft coupling
$\lambda\gg 1$, only protected (chiral) operators have finite
dimensions $\Delta \ll \lambda^{1/4}$ and that the lowest-dimensional
fermionic operator has dimension $2{+}\frac12=\frac52$
(being the supersymmetry variation of a dimension-2 primary).
Similarly, the ${\cal N}=1$ theory dual to IIB string theory on
$\rm{AdS}_5\times {\rm T}^{11}$ \cite{klebanovwitten}
is known to contain no fermionic operator of dimension less than 2
\cite{t11spectrum}.
Thus, in these theories, supersymmetry violations can only
be seen at the level of $\sim E^{-3}$ or $\sim E^{-\frac52}$ corrections,
respectively.
A discussion of more general strongly
coupled theories will not be attempted here.



\section{Discussion}

In this letter we have shown that supersymmetry is a generic property
of the effective theories which describe
high-energy correlators in supersymmetric plasmas.
These correlators included self-energies at high energies
on the light-cone as well as far away from it (large virtuality).

For all correlators (except for the more tractable
deeply virtual correlators, treated in section \ref{sec:ope})
our analysis has been limited to the leading nontrivial
order in the relevant coupling constant expansion
(at both weak and strong coupling).
Unfortunately, this seems like a major limitation of this work:
it makes it hard to decide whether our findings highlight
general properties of supersymmetric theories, or whether
they are mere artefacts of these extreme limits.
In this respect it would be interesting to know the structure of
higher order corrections, at both weak and strong coupling.

We have found that thermal supersymmetry violations in all correlators
are suppressed by a power of the energy $E^{-n}$
(relative to the vacuum correlators),
with $n$ \emph{strictly greater than $2$}.
(Violations with $n=2$ were observed
in sections \ref{sec:masses} and \ref{sec:ope}
due to nonvanishing $D$-term of $F$-term expectation values,
but we do not regard these as specifically thermal effects.
These would have identical supersymmetry-breaking effects
in vacuum correlators.)

We find pleasing that such a simple uniform bound holds.
This makes one wonder whether our findings could be elevated
to some sort of theorem, valid independently of perturbation theory,
though we have no concrete proposal to make along these lines.

\section{Acknowledgements}

I would like to thank Guy~D.~Moore and K.~Dasgupta for useful
discussions and comments on a draft of this paper,
and O.~Saremi for helpful advice on fermions in AdS/CFT.
This work was supported in part by
the Natural Sciences and Engineering Research Council of Canada.

\end{document}